\documentclass[11pt, titlepage]{article}
\usepackage{amscd}
\usepackage{amssymb}
\usepackage{amsmath}
\usepackage{stmaryrd}
\usepackage{mathrsfs}

\newcommand{\zot}{$\mathbb{Z}$ot}

\title{A User's Guide to \zot{}}

\author{Matteo Pradella\\
\\
CNR IEIIT, Milano, Italy\\
pradella@elet.polimi.it\\
http://home.dei.polimi.it/pradella/
}

\date{December 2009}

\begin{document}

\maketitle
\newpage

\pagenumbering{roman}
\tableofcontents
\newpage

\pagenumbering{arabic}

\section{Overview}

\zot{} is an agile and easily extendible bounded model checker, which can
be downloaded at http://home.\-dei.\-polimi.\-it/\-pradella/.

The tool supports different logic languages through a multi-layered
approach: its core uses PLTL, and on top of it a decidable predicative
fragment of TRIO \cite{GMM90} is defined. An interesting feature of
\zot{} is its ability to support different encodings of temporal logic as
SAT problems by means of plug-ins. This approach encourages
experimentation, as plug-ins are expected to be quite simple, compact
(usually around 500 lines of code), easily modifiable, and
extendible. At the moment, a variant of the eventuality encoding
presented in \cite{BH+06} is supported, (approximated) dense-time MTL
\cite{FPR07}, and a bi-infinite encoding \cite{PMS07}, \cite{PMS-ICTAC08}.

\zot{} offers three basic usage modalities:
\begin{enumerate}
\item {\em Bounded satisfiability checking (BSC)}: given as input a
  specification formula, the tool returns a (possibly empty) history
  (i.e., an execution trace of the specified system) which satisfies
  the specification. An empty history means that it is impossible to
  satisfy the specification.

\item {\em Bounded model checking (BMC)}: given as input an
  operational model of the system, the tool returns a (possibly empty)
  history (i.e., an execution trace of the specified system) which
  satisfies it.

\item {\em History checking and completion (HCC)}: The input
  file can also contain a partial (or complete) history
  $H$. In this case, if $H$ complies with the specification, then a
  completed version of $H$ is returned as output, otherwise the output
  is empty.
\end{enumerate}

The provided output histories have temporal length $\le k$,
the bound given by the user, but may represent infinite behaviors
thanks to the loop selector variables, marking the start of the
periodic sections of the history. The BSC/BMC modalities can be used to
check if a property {\em prop} of the given specification {\em
spec} holds over every periodic behavior with period $\le k$. In
this case, the input file contains $\mathrm{spec} \land \neg
\mathrm{prop}$, and, if $\mathrm{prop}$ indeed holds, then the
output history is empty. If this is not the case, the output
history is a counterexample, explaining why
$\mathrm{prop}$ does not hold.

\newpage

\section{Installation}

\zot{}'s core is written in Common Lisp (with ASDF packaging
http://www.\-cliki.\-net/asdf).  It can be used under Linux, Windows,
or MacOS X, but has been tested only under Linux and Windows XP, using the
following
Common Lisps\footnote{SBCL and CMUCL are usually the fastest
implementations, for running Zot.}: 
\begin{itemize}
	\item SBCL (http://www.sbcl.org), 
	\item CLISP (http://clisp.\-cons.\-org),
	\item CMUCL (http://www.cons.\-org/\-cmucl/),
	\item ABCL (http://common-\-lisp.net/\-project/\-armedbear/),
	\item Clozure CL (http://www.clozure.com/\-clozurecl.html),
\end{itemize}
This approach makes \zot{} an open system, as
it uses Common Lisp also as internal scripting language of the tool,
both to define complex verification activities, and to add new
constructs and languages on top of the existing ones.

Typically, to install \zot{} in a Debian system (or Ubuntu), the user
must install a Common Lisp (e.g. one of the packages \verb+clisp+,
\verb+sbcl+, \verb+cmucl+, \dots),
and the \verb+common-lisp-controller+ package. To perform a system-wide
install of the \zot{} packages, just put symbolic links to its .ads files
in the /usr/share/common-lisp/systems/ directory. Note that it is
possible to avoid a system-wide installation, but in this case the
user has to work inside the main \zot{} directory.

\zot{} works with external SAT-solvers.
The supported SAT-solvers are MiniSat (default) \cite{minisat}, 
MiraXT \cite{miraxt}, PicoSAT \cite{picosat}, and zChaff \cite{Chaff01}. \zot{} assumes
that executable files called 
\verb+minisat+, 
\verb+MiraXTSimp+ (optional),
\verb+picosat+ (optional),
\verb+zchaff+ (optional), 
are system-wide installed.

A pre-packaged all-inclusive version for Windows ({\em WinZot}, based
on Cygwin-compiled binaries and SBCL) is available from the author.

All \zot{}'s components are available as open source software (GPL v2).

\newpage

\section{Languages}

Being an open system, \zot{} supports different languages. At present,
the main native language is PLTL (linear temporal logic with future
and past operators). The other main layer based on PLTL is the metric
temporal logic TRIO.

\zot{} scripts are written in Common Lisp, so a basic knowledge of the
language is required. It is very easy to find online a lot of
tutorials and short
presentations\footnote{e.g. http://gigamonkeys.com/book/ is a good and
freely available text.}.

\subsection{PLTL}

\noindent {\bf Propositional operators} are written as: \verb+&&+
(and), \verb+||+ (or), \verb+!!+ (not). \\

\noindent {\bf Predicates and propositional letters}
e.g., proposition Q is written (-P- Q); 
predicate Pred(1,2) is written as (-P- Pred 1 2).\\

\noindent {\bf Quantifications} 
$\exists t \in \{One, Two\} : Formula(t)$ 
is written\\ \verb+(-E- t '(One Two) Formula(t))+.
-A- is the universal quantifier.

Term comparisons and conditions are available through Common Lisp
(e.g. eql, equal, $<$, $<=$, and, or, not, \ldots)\\

\noindent {\bf Temporal operators} The following temporal operators
are supported: 
\verb+until+,
\verb+since+,
\verb+release+,
\verb+trigger+,
\verb+next+,
\verb+yesterday+,
\verb+zeta+. 
The last one is the dual of \verb+yesterday+, and is used only in the
mono-infinite semantics.

For the semantics of these operators, see e.g. \cite{BH+06}
(which describes the implementation of the mono-infinite
encoding in details).

\subsection{TRIO}\label{trio}

\zot{} was originally born as a satisfiability checker for the TRIO
metric temporal logic \cite{GMM90}.

The list of supported operators (and their correct ``\zot{} spelling'')
is the following:
\begin{verbatim}
   dist
   futr
   past
   lasts    lasts_ee    lasts_ie    lasts_ei    lasts_ii
   lasted   lasted_ee   lasted_ie   lasted_ei   lasted_ii
   withinf  withinf_ee  withinf_ie  withinf_ei  withinf_ii 
   withinp  withinp_ee  withinp_ie  withinp_ei  withinp_ii 
   lasttime lasttime_ee lasttime_ie lasttime_ei lasttime_ii
   nexttime nexttime_ee nexttime_ie nexttime_ei nexttime_ii
   somf     somf_e      somf_i      som
   somp     somp_e      somp_i
   alwf     alwf_e      alwf_i      alw
   alwp     alwp_e      alwp_i
   until    until_ie    until_ee    until_ii    until_ei
   since    since_ie    since_ee    since_ii    since_ei
\end{verbatim}

Bounded version of since and until are written as:
\begin{verbatim}
    (until_ie_<=_<= t1 t2 A B) 
    B will be true at t instants in the future with t1<=t<=t2
    (until_ie_>= t1 A B) 
    B will be true at t instants in the future with t>=t1 
    since_ie_<=_<=
    since_ie_>=
\end{verbatim}

Caveat emptor! The default \verb+until+ is PLTL's (which is
usually called \verb+until_ie+ in TRIO).  For example, the following
model satisfies \verb+(until A B)+ at 0:
\begin{verbatim}
    0                    B
    ----------------------
    AAAAAAAAAAAAAAAAAAAAA
\end{verbatim}
B may appear at 0.

\medskip

\noindent {\bf For MTL users}:
\begin{enumerate}
	\item $\lozenge_{= t} A$ (or $\Box_{= t} A)$) is written \verb+(futr (-P- A) t)+;
	\item $\Box_{\le t} A$ is written \verb+(lasts (-P- A) t)+;
	\item $\lozenge_{\le t} A$ is written \verb+(withinf (-P- A) t)+;

	\item $\blacklozenge_{= t} A$ (or $\blacksquare_{= t} A)$) is written
		\verb+(past (-P- A) t)+;
	\item $\blacksquare_{\le t} A$ is written \verb+(lasted (-P- A) t)+;
	\item $\blacklozenge_{\le t} A$ is written \verb+(withinp (-P- A) t)+;
\end{enumerate}
with $t > 0$.

\subsection{Operational constructs}\label{op}

\zot{} offers some simple facilities to describe operational systems.

\noindent
\begin{verbatim}
(define-item  <varname> <domain>)
\end{verbatim}
is used to define variables \`a la Von Neumann over finite domains
(e.g. counters).

\noindent
\begin{verbatim}
(define-array <varname> <index-domain> <domain>)
\end{verbatim}
is used to define mono-dimensional arrays.\\

Example usage:
\begin{verbatim}
(define-item  cont (loop for i from 0 to 9 collect i))
(define-array arr  (loop for i from 0 to 9 collect i) 
                   '(on off unknown))
\end{verbatim}
In the spec, the user can e.g. write \verb+(cont= 6)+;
\verb+(arr= 6 'off)+.

Caveat: both define-item and define-array have side effects. It is
therefore wrong to ``define-items'' after a zot main procedure call, since
successive calls may work with spurious constraints. It is therefore
recommended to perform \verb+(clean-up)+ before defining items or arrays.

Typically, to define an operational model means to constraint
operational variables and arrays. This can be done either by using
simple next-time formulae, i.e. containing only the \verb+next+
temporal operator, or by using the two dual constructs  
\verb+and-case+ and \verb+or-case+ \cite{PMS-ASE08}.

To give the reader an idea of their semantics, here is an
automatic translation made by \zot{} on two simple examples.
\begin{verbatim}
 (and-case (x '(1 2) y '(3 4))
           ((-P- P x) (-P- Q x))
           ((-P- R y) (-P- R1 y))
           (else (-P- R2 x))) 
\end{verbatim}
           
expands to

\begin{verbatim}
 (-A- X '(1 2)
     (-A- Y '(3 4)
      (&& (-> (-P- R Y) (-P- R1 Y)) (-> (-P- P X) (-P- Q X))
       (-> (&& (!! (-P- R Y)) (!! (-P- P X))) (-P- R2 X)))))
\end{verbatim}

and
           
\begin{verbatim}
 (or-case (x '(1 2) y '(3 4))
           ((-P- P x) (-P- Q x))
           ((-P- R y) (-P- R1 y))
           (else (-P- R2 x)))
\end{verbatim}
           
expands to

\begin{verbatim}
 (-E- X '(1 2)
     (-E- Y '(3 4)
      (|| (&& (-P- R Y) (-P- R1 Y)) (&& (-P- P X) (-P- Q X))
       (&& (!! (-P- R Y)) (!! (-P- P X)) (-P- R2 X)))))
\end{verbatim}

\subsection{MTL}\label{ap}

There is an experimental plug-in (called {\em ap-zot} for using a
variant of dense-time MTL through approximation (see \cite{FPR07}, and
\cite{FPR07-rep}).

Here is a list of the time operator defined in ap-zot.
\begin{verbatim}
    until-b   until-b-v   until-b-^
    since-b   since-b-v   since-b-^
    release-b release-b-^ release-b-v
    trigger-b trigger-b-^ trigger-b-v
    
    until-b-inf   until-b-v-inf   until-b-^-inf
    since-b-inf   since-b-v-inf   since-b-^-inf
    release-b-inf release-b-^-inf release-b-v-inf
    trigger-b-inf trigger-b-^-inf trigger-b-v-inf

    diamond   diamond-inf
    diamond-p diamond-inf-p
    box       box-inf
    box-p     box-inf-p
\end{verbatim}

The plug-in offers the following operations
\begin{verbatim}
   normalize
   basicize
   compute-granularity
   over-approximation
   under-approximation
   nth-divisor
\end{verbatim}

To compute over- and under-approximations, an axiom must be prepared
through the two functions {\em basicize} and {\em normalize}\\ (e.g. 
with \verb+(setf ax1 (normalize (basicize ax1)))+).

The two functions {\em over-approximation} and {\em
  under-approximation} are used to compute the approximated formulae,
while {\em compute-granularity} is used to set the $\rho$ parameter
(see \cite{FPR07} for details).

The interested reader may find a complete example in 
\verb+coffee.lisp+.

\subsection{Timed Automata}\label{ta}

Timed Automata (TA) are supported through a {\em very} experimental
plug-in called {\em ta-zot} (see \cite{FPR08-rep}, \cite{fpr-icfem08}), which is based on
the approximations offered by {\em ap-zot}.

First, here is a list of the added operators, and approximations
procedures:
\begin{verbatim}
	   white-tri
	   white-tri/3
	   black-tri
	   black-tri/3

	   timed-automaton-under-formula
	   timed-automaton-over-formula

	   timed-automata-under-formula
	   timed-automata-over-formula
\end{verbatim}

Here is the main data structure used to represent TA's, together with
its interface:
\begin{verbatim}
   (defstruct timed-automaton
     alphabet
     states
     initial-states
     clocks)
   
   (defgeneric add-trans (autom from to lamb constr))
   (defgeneric add-label (autom state list-of-symbols))
   (defgeneric alpha (autom state))
   (defgeneric get-trans-from-states (autom from to))
   (defgeneric all-connected-pairs (autom))
   (defgeneric all-unconnected-pairs (autom))
   (defgeneric get-all-trans (autom))
   (defgeneric get-trans-from-clock-reset (autom clock))
\end{verbatim}

The interested reader may find a complete example in
\begin{verbatim}
    trans_prot.lisp.
\end{verbatim}

\newpage

\section{Usage}

\subsection{SAT-solvers}

The supported SAT-solvers are MiniSat \cite{minisat} (which is used by
default), MiraXT \cite{miraxt}, and zChaff \cite{Chaff01}.

To use the zChaff SAT-solver, the user has to set the *zot-solver*
parameter.
For example:
\begin{verbatim}
   (setq sat-interface:*zot-solver* :zchaff)
\end{verbatim}

MiraXT is a multi-threaded solver, so to use it we also have to choose
the maximum number of threads that it will use:
\begin{verbatim}
   (setf sat-interface:*zot-solver* :miraxt)
   (setf sat-interface:*n-threads* 3)
\end{verbatim}

\subsection{Model Checking}

To perform Bounded Model Checking, the user must provide the model through
as argument :transitions. Important: every variable used must be declared
implicitly by e.g. an initialization formula as the second argument of \zot{}.


Here is a simple example: mutex3 (a simple mutual exclusion protocol
with three processes).

The first part is used to load the mono-infinite plug-in, and defines
the used variables.
The first line loads the mono-infinite plug-in, called {\em
  eezot}. ({\em bezot} is the bi-infinite one.)
\begin{verbatim}
(asdf:operate 'asdf:load-op 'eezot)
(use-package  :trio-utils)

(defvar state-d '(N T C))
(defvar turn-d  '(1 2 3))


(define-array state turn-d state-d)
(define-item  turn  turn-d)

(defconstant decl ; optional declarations, just for checking usage
  (append
   (loop for x in state-d append
         (loop for y in turn-d collect (state= y x)))
   (loop for x in turn-d collect (turn= x))))
\end{verbatim}

Then, we define the system initialization and transitions:
\begin{verbatim}
(defvar init   ; system initialization (at 0)
  (&& (-A- x turn-d (state= x 'N))
      (turn= 1)))

(defvar trans  ; list of model constraints
  (list
   (-A- p turn-d 
        (or-case (x state-d)
                 ((state= p 'N)         
                  (next (state= p 'T)))
                 
                 ((&& (state= p 'T)
                      (|| (-A- p1 turn-d (-> (not (equal p p1))
                                             (state= p1 'N)))
                       (turn= p)))      
                  (next (state= p 'C)))
                 
                 ((state= p 'C)
                  (next (state= p 'N)))
                 
                 (else            
                  (&& (state= p x)
                      (next (state= p x))))))

   (or-case (x turn-d)  ; -- schedule --

             ((&& (state= 1 'N) (state= 2 'T) (state= 3 'N)) 
              (next (turn= 2)))
             ((&& (state= 1 'T) (state= 1 'N) (state= 3 'N)) 
              (next (turn= 1)))
             ((&& (state= 1 'N) (state= 1 'N) (state= 3 'T)) 
              (next (turn= 3)))

   
           ; --- random choice policy ---
             ((&& (state= 1 'T)(state= 2 'T))   
              (next (|| (turn= 1)(turn= 2))))
             ((&& (state= 1 'T)(state= 3 'T))   
              (next (|| (turn= 1)(turn= 3))))
             ((&& (state= 2 'T)(state= 3 'T))   
              (next (|| (turn= 2)(turn= 3))))

             (else  
              (&& (turn= x) (next (turn= x)))))))
\end{verbatim}
As the reader may see, the transitions are defined as a list of
constraints, which must hold on every instant of the time domain.

We then write a simple property we wish to check on the system:
\begin{verbatim}
(defvar spec
    (alw
      (&& 
        (-> (turn= 1) (somf (|| (turn= 2)(turn= 3)))) 
        (-> (turn= 2) (somf (|| (turn= 1)(turn= 3)))) 
        (-> (turn= 3) (somf (|| (turn= 1)(turn= 2)))))))
\end{verbatim}

The main procedure is called {\em zot}, and has two arguments: the
time bound and the formula to be satisfied (plus some optional switches,
e.g. :transitions, :declarations, :loop-free).

To check if spec-0 holds for a time bound of 30, we perform:
\begin{verbatim}
(eezot:zot 30               ; time bound
      (&& (yesterday init) ; initialization (init must hold at 0)
          (!! spec))       ; (negated) property
      :transitions trans   ; list of model constraints
      :declarations decl   ; (optional) declarations
      )
\end{verbatim}
UNSAT means that the desired property holds. If the output is SAT,
then {\em spec} does not hold and \zot{} returns a counter-example.

\subsection{Completeness}

A switch of the {\em zot} procedure (:loop-free, \verb+nil+ by default) is
used to check completeness. In the previous example, we can check
completeness by performing:
\begin{verbatim}
(eezot:zot 30            ; time bound
     (yesterday init)   ; initialization (init must hold at 0)
     :transitions trans ; list of model constraints
     :declarations decl ; (optional) declarations
     :loop-free t       ; check completeness
     )
\end{verbatim}

UNSAT means that the completeness bound is reached.\\

The {\em zot} procedure returns \verb+t+ if the spec is satisfiable,
\verb+nil+ otherwise. So, it is possible to write a loop to actually
find the completeness bound, e.g.:
\newpage
\begin{verbatim}
(format t "Found: ~s~%"
        (loop for bound from 2 unless
             (eezot:zot bound    
                       (yesterday init)  
                       :transitions trans
                       :declarations decl
                       :loop-free t      
                       )
           return bound))
\end{verbatim}

\subsection{Satisfiability Checking}

Let us now consider a simple example to show how satisfiability
checking can be performed with \zot{}.

The first line loads the bi-inifinite plug-in.
\begin{verbatim}
(asdf:operate 'asdf:load-op 'bezot)
(use-package  :trio-utils)
\end{verbatim}

We then define the timed lamp spec:
\begin{verbatim}
(defconstant delta 5)

; Alphabet
; on:  the "on"  button is pressed
; off: the "off" button is pressed
; L:   the light is on

(defconstant init 
  (&& (!! (|| (-P- on)(-P- off)(-P- L)))))

(defconstant the-lamp
  (alw (&& 
         (<->
          (-P- L)
          (|| (yesterday (-P- on))
              (-E- x (loop for i from 2 to delta collect i)
                   (&& (past (-P- on) x)
                       (!! (withinP_ee (-P- off) x))))))
         (!! (&& (-P- on) (-P- off))))))
\end{verbatim}

To obtain a history compatible with the spec, we perform:
\begin{verbatim}
(bezot:zot  10 
    (&& init the-lamp))
\end{verbatim}

This is an example history generated by \zot{}, where **LOOP**, and
**POOL** are the loop selector variables (**POOL** towards the
past, **LOOP** towards the future):
\begin{verbatim}

    ------ time 0 ------
    
    ------ time 1 ------
      **LOOP**
      ON
    
    ------ time 2 ------
      ON
      L
    
    ------ time 3 ------
      ON
      L
    
    ------ time 4 ------
      OFF
      L
    
    ------ time 5 ------
      OFF
    
    ------ time 6 ------
      OFF
    
    ------ time 7 ------
      OFF
    
    ------ time 8 ------
      OFF
    
    ------ time 9 ------
      **POOL**
      OFF
    
    ------ time 10 ------
    
    ------ end ------
\end{verbatim}

\subsection{Temporary data}
\zot{} uses four files to save temporary data during the
verification activity:
\begin{verbatim}
    1) output.cnf.txt
    2) output.sat.txt
    3) output.hist.txt
\end{verbatim}

(1) contains the resulting boolean formula of the system (in the
standard DIMACS CNF format); (2) is the output of the SAT-solver; (3)
is the resulting trace of the system (e.g. a TRIO history).

\newpage

\section{Architecture}

\zot{}'s architecture is based on a PLTL-to-SAT core, which interacts
with the ``outside world'' through a TRIO-based interface and
different plug-ins.  The core itself is structured as a plug-in, so
that different encodings can be defined and used.

More recently (May 2009), we added two plugins to \zot{}, natively
supporting metric operators (like {\em lasts, withinf}). These native
metric plugins are called {\em meezot} (mono-infinite), and {\em mbezot}.
Their usage is exactly the same as {\em eezot} and {\em bezot} \cite{PMS09}.

\subsection{PLTL-to-SAT encodings}

As said before, \zot{}'s core is based on encoding PLTL into SAT. At
present two main encodings are available in the standard distribution:
{\em eezot}, which is a standard eventuality-based encoding on a
mono-infinite time domain ($\mathbb{N}$, see e.g. \cite{BH+06}), 
and the bi-infinite one, {\em bezot} \cite{PMS07} on $\mathbb{Z}$. 

The two encodings are packaged (as asdf systems) in the following files:
\begin{verbatim}
    eezot.lisp   eezot.asd
    bezot.lisp  bezot.asd 
\end{verbatim}

The file \verb+kripke.lisp+ contains the basic data structure and the
definition of the generics.\footnote{{\em kripke} does not actually
  contain a Kripke structure - names of data structures and generics
  come from previous, forsaken incarnations of the tool-set.}

\begin{verbatim}
(defclass kripke ()
  (; time bound i.e. [0..k]
   (the-k       :accessor kripke-k) 

   ; number of used prop. variables
   (numvar      :accessor kripke-numvar)

   ; formula -> integer data structure (hash-table)
   (the-list    :accessor kripke-list)

   ; integer -> formula data structure (hash-table)
   (the-back    :accessor kripke-back)
     
   ; list of propositional letters
   (sf-prop     :accessor kripke-prop)

   ; list of used boolean subformulae
   (sf-bool     :accessor kripke-bool)

   ; list of used future-tense subf.
   (sf-futr     :accessor kripke-futr)

   ; list of used past-tense subf.
   (sf-past     :accessor kripke-past)

   ; n. of props used in the encoding
   (max-prop    :accessor kripke-maximum)

   ; resulting SAT formula
   (the-formula :accessor kripke-formula)))
\end{verbatim}

There is also an old variant of {\em eezot}, called {\em ezot},
which supports virtual unrollings (as presented in \cite{BH+06},
usually called $\delta$), so its data structure is extended through
inheritance.  The user may change the default behavior (i.e. $\delta =
0$), by setting ezot:*FIXED-DELTA* to nil, which tells eezot to
actually compute $\delta$, or (s)he may change to set it to a fixed
meaningful value.

The {\em call} generic translates a formula/proposition and a time
instant into an integer (the SAT-solver proposition); {\em self} must
be an instance of kripke (or of a subclass).
\begin{verbatim}
  (defgeneric call (self obj the-time &rest other-stuff))
\end{verbatim}

The {\em back-call} generic is used to translate an integer in
$[0..k]$ into the corresponding subformula; self must be an instance of
kripke (or of a subclass).
\begin{verbatim}
  (defgeneric back-call (self x))
  (defgeneric back-call-time (self x))
\end{verbatim}

\subsection{Main Interface}

There are two interfaces: 
\begin{verbatim}
    sat-interface.lisp
\end{verbatim}
the first one is with the
SAT-solver, and it is used to send the output of the PLTL encoding to
it; then, to parse its output and get a counter-example, if any.

The other one,
\begin{verbatim} 
    trio-utils.lisp
\end{verbatim}
is the basic interface with the user, and is based on TRIO (see
Section \ref{trio}) augmented with the operational constructs covered
in Section \ref{op}.

\newpage
\subsection{Other modules and plug-ins}

At present just {\em ap-zot} and {\em ta-zot} are available. Please
refer to Sections \ref{ap}, \ref{ta}, and the related papers.

The two plug-ins are implemented and packaged (as asdf systems) in 
\begin{verbatim} 
    ap-zot.lisp  ap-zot.asd
    ta-zot.lisp  ta-zot.asd
\end{verbatim}

{\em ta-zot} is based on {\em ap-zot}, which uses TRIO as underlying
language (through the {\em trio-utils} interface).

\subsection*{Acknowledgments}
I thank the following people: Stefano Riboni for his work on the CNF translator; Davide
Casiraghi for the metric plugins (meezot and mbezot).

\newpage
\bibliographystyle{abbrv}
\bibliography{triobib,automatabib}

\end{document}